\documentclass[aps,prb,twocolumn,superscriptaddress]{revtex4-2}
\usepackage{graphicx}
\usepackage{epstopdf}
\usepackage{amsmath}
\begin{document}
\title{Phonon Hall effect with first-principles calculations}
\author{Kangtai Sun}
\thanks{\href{mailto:E0212212@u.nus.edu} {E0212212@u.nus.edu}}
\affiliation{Department of Physics, National University of Singapore, Singapore 117551, Republic of Singapore}
\author{Zhibin Gao}
\thanks{\href{mailto:zhibin.gao@xjtu.edu.cn} {zhibin.gao@xjtu.edu.cn}}
\affiliation{Department of Physics, National University of Singapore, Singapore 117551, Republic of Singapore}
\affiliation{State Key Laboratory for Mechanical Behavior of Materials, Xi'an Jiaotong University, Xi'an 710049, China}
\author{Jian-Sheng Wang}
\affiliation{Department of Physics, National University of Singapore, Singapore 117551, Republic of Singapore}
\revised{17 May 2021}

\begin{abstract}
Phonon Hall effect (PHE) has attracted a lot of attention in recent years with many theoretical and experimental explorations published. While experiments work on complicated materials, theoretical studies are still hovering around the phenomenon-based models. Moreover, previous microscopic theory was found unable to explain large thermal Hall conductivity obtained by experiments in strontium titanate (STO). Therefore, as a first attempt to bridge this gap, we implement first-principles calculations to explore the PHE in real materials. Our work provides a new benchmark of the PHE in sodium chloride (NaCl) under a large external magnetic field. Moreover, we demonstrate our results in barium titanate (BTO), and discuss the results in STO in detail about their deviation from experiments. As possible future directions, we further propose that the inner electronic Berry curvature or cubic potential plays important roles in the PHE in STO. 
\end{abstract}
\maketitle

\section{Introduction}
PHE, as a phonon analogue to the quantum Hall effect of electrons, has been a rather intriguing area since its first observation in 2005 \cite{strohm2005phenomenological}. In the past decade, several theoretical explanations and mechanisms have been proposed \cite{PhysRevB.80.012301, sheng2006theory, kagan2008anomalous, PhysRevLett.105.225901, PhysRevB.86.104305, saito2019berry, PhysRevLett.123.167202, PhysRevB.102.134311}. Currently, the most successful microscopic theory was developed by Qin \emph{et~al.} in which the PHE is related to the topological properties of the phononic structure \cite{PhysRevB.86.104305}. However, with more experiments published, it is evident that we have not reached the end of the story yet. In 2020, an experimental group found a large PHE in STO, and they thought it can be explained by Qin's theory \cite{li2020phonon}. However, subsequently, a theoretical group pointed out that Qin's theory cannot explain the large value in experiments and they used Boltzmann transport theory to successfully predict the ratio between the longitudinal thermal conductivity and the phonon Hall conductivity \cite{chen2020enhanced}. Furthermore, another experimental work found that if the $^{16}$O in STO is replaced with its isotope $^{18}$O, the phonon Hall conductivity will become two orders of magnitude smaller \cite{sim126sizable}. This is a very bizarre behavior challenging all current theories. The authors concluded that the PHE in STO with $^{16}$O is more like an enhancement compared with SrTi$^{18}$O$_3$, and therefore they attributed the reason most likely to the behavior of the transverse optical phonon modes in STO at low temperature. All these recent experiments are performed on complex materials, therefore, it is difficult to understand them with simplified models, and more accurate and persuasive first-principles calculations are needed. 

Usually, harmonic assumption is made in first-principles calculations for phonon properties like phonon dispersion. However, in some highly anharmonic materials, harmonic terms alone will produce imaginary phonon frequencies, and they cannot explain those phenomena such as the thermal expansion, temperature-dependent phonon dispersion, and some phase transitions. Therefore, beyond harmonicity is a natural requirement to explore the PHE in complex materials like STO, and it was argued in a similar perovskite, BTO, that the anharmonic soft phonon modes will result in a large dielectric constant \cite{wang2020phonon} which could act as a magnifier of the PHE \cite{chen2020enhanced}. Based on this understanding, anharmonicity should play an important role in the PHE in real materials. In recent years, many packages based on first-principles calculations have been developed to calculate anharmonic properties in solids such as SCAILD, ALAMODE, and TDEP \cite{souvatzis2008entropy, tadano2015self, hellman2013temperature}. With the help of these packages, it is feasible for us to study the PHE in real materials, which could deepen our understandings in this area.

The paper is organized as follows. In section II, we describe the self-consistent phonon calculation, which is the first step to calculate the PHE. In section III, we introduce the general PHE theory and discuss how to apply it to real materials utilizing the results obtained by the self-consistent phonon calculation. In section IV, we present our results for NaCl and BTO, and discuss the situation in STO. In section V, we draw conclusions of our work and propose that there is still a lot of future work required to fully understand the PHE. We also provide an appendix with some key details.

\section{Anharmonic self-consistent phonon calculation for soft phonon modes}
There are currently three approaches to handle the anharmonicity: density functional perturbation theory \cite{baroni2001phonons, esfarjani2008method}, \textit{ab initio} molecular dynamics (AIMD) \cite{sun2014dynamic, hellman2013temperature}, and self-consistent phonon (SCPH) theory \cite{werthamer1970self, souvatzis2008entropy, errea2014anharmonic, tadano2015self}. Perturbation theory is only valid for weak anharmonicity, while AIMD is a nonperturbative approach. However, since AIMD is based on the time-dependent Schr\"{o}dinger equation for all particles approximately \cite{PhysRevLett.55.2471}, it cannot include zero-point vibration which is significant at low temperature. SCPH provides another choice to address anharmonicity nonperturbatively considering the quantum effect. Therefore, in this study, we focus on the SCPH approach, and borrow the ALAMODE package developed by Tadano and Shinji \cite{tadano2015self}. In this section, we briefly introduce the SCPH theory.

A general Hamiltonian with the third- and the fourth-order Taylor expansion of the potential can be described as follows:
\begin{equation}
\begin{aligned}
\hat{H} &= \frac{1}{2}\sum\limits_ip_i^2 + \frac{1}{2}u^TKu \\
            &+ \frac{1}{3}\sum\limits_{ijk}\Gamma_{ijk}u_iu_ju_k + \frac{1}{4}\sum\limits_{ijkl}T_{ijkl}u_iu_ju_ku_l,
\end{aligned}
\end{equation}
where $u_i\equiv\sqrt{M_i}x_i$, $M_i$ and $x_i$ are mass and displacement of the $i$-th degree of freedom, respectively. Although the third-order term has important contribution for most anharmonic behaviors like thermal expansion and phonon lifetime, the fourth-order term is also important, especially for the soft phonon modes. Moreover, the fourth-order is simpler than the third if we apply a mean-field approximation by replacing $u^4$ with $\langle u^2\rangle u^2$. With this approximation, the problem goes back to a quadratic one with the effective force constants being determined self-consistently. Therefore, in this paper, we only focus on the fourth-order correction. By an equation of motion method \cite{xu2008nonequilibrium}, the non-equilibrium Green's function (NEGF) satisfies:
\begin{equation}
\begin{aligned}
G(1, 2) &= G_0(1, 2)\\ 
            &+ \int d1^{\prime}d2^{\prime}d3d4G_0(1, 1^{\prime})T(1^{\prime}, 2^{\prime}, 3, 4)G(2^{\prime}, 3, 4, 2),
\end{aligned}
\end{equation}
where $G(1,2)=-\frac{i}{\hbar}\langle \hat{T}u(1)u(2)\rangle$, $G(1,2,3,4)=-\frac{i}{\hbar}\langle \hat{T}u(1)u(2)u(3)u(4)\rangle$, $\hat{T}$ is the contour order operator, $G_0(1,2)$ is the non-interacting version of $G(1,2)$, and numbers represent the combination of $(jt)$. To close this equation, we need to apply a mean-field approximation:
\begin{equation}
\begin{aligned}
G(1,2,3,4) &\approx i\hbar\Big[G(1,2)G(3,4) + G(1,3)G(2,4) \\
               &+ G(1,4)G(2,3)\Big].
\end{aligned}
\end{equation}
Then we can work out the effective force constant matrix: 
\begin{equation}
K^e = K + \Sigma,
\end{equation}
where we define
\begin{equation}
\Sigma_{ij} =3\sum\limits_{kl}T_{ijkl}\langle u_ku_l\rangle .
\end{equation}
The ingredient we need in the PHE is the dynamic matrix, and therefore we need to transform the equation into mode space, which is:
\begin{equation}\label{eq:dynamic}
\begin{aligned}
D_{nn^{\prime}}(\boldsymbol{q}) &= \omega_n(\boldsymbol{q})^2\delta_{nn^{\prime}} \\
&+ 3\sum\limits_{mm^{\prime}\boldsymbol{q^{\prime}}}T_{nn^{\prime}mm^{\prime}}(\boldsymbol{q},\boldsymbol{q^{\prime}})\langle Q_m(\boldsymbol{q^{\prime}})Q_{m^{\prime}}(\boldsymbol{q^{\prime}})^*\rangle, 
\end{aligned}
\end{equation}
where $Q$ represents normal modes, $n$ and $m$ are indices for normal modes, $\boldsymbol{q}$ and $\boldsymbol{q^{\prime}}$ are lattice momentum, and $\omega_n$ is the eigenfrequency. This equation should be solved self-consistently. In 2015, Tadano and Shinji have already discussed details within their ALAMODE package \cite{tadano2015self}. Therefore, we utilize their package to calculate the dynamic matrix for real materials.

\section{Phonon Hall effect theory}
Currently, the widely accepted general theory for the PHE was proposed by Qin $\emph{et~al.}$ in 2012 \cite{PhysRevB.86.104305}. This theory introduces an effective vector potential to explain the PHE. We describe a harmonic phonon system in the reciprocal space with a Hamiltonian $\hat{H}=\frac{1}{2}\sum_{\boldsymbol{q}}y_{\boldsymbol{q}}^{\dagger}H(\boldsymbol{q})y_{\boldsymbol{q}}$, where $H_{\boldsymbol{q}}=\mathrm{diag}\{D_{\boldsymbol{q}}, I\}$ with $D_{\boldsymbol{q}}$ being the dynamic matrix, $y_{\boldsymbol{q}}=(\boldsymbol{u}_{\boldsymbol{q}}, \boldsymbol{v}_{\boldsymbol{q}})^T$, $u_{\boldsymbol{q}}$ is the displacement vector multiplied by the associated mass, and $\boldsymbol{v}_{\boldsymbol{q}}=\dot{\boldsymbol{u}}_{\boldsymbol{q}}$ is the corresponding velocity vector. The $y_{\boldsymbol{q}}$ satisfies the commutation relation:
\begin{equation}
\begin{aligned}
&[y_{\boldsymbol{q}}, y_{\boldsymbol{q}^{\prime}}^{\dagger}] = i\hbar J(\boldsymbol{q})\delta_{\boldsymbol{q}\boldsymbol{q}^{\prime}},\\
&J(\boldsymbol{q})=\begin{pmatrix} 0 &I\\ -I &-2A_{\boldsymbol{q}} \end{pmatrix},
\end{aligned}
\end{equation}
where $A_{\boldsymbol{q}}$ is an anti-Hermitian matrix. Assuming $y_{\boldsymbol{q}}=\psi_{\boldsymbol{q}}e^{-i\omega_{\boldsymbol{q}}t}$, we obtain the following eigen equation \cite{PhysRevB.102.134311}:
\begin{equation}\label{eq:omega}
\omega_{\boldsymbol{q}i}\psi_{\boldsymbol{q}i} = \begin{pmatrix} 0 &iI\\ -iD_{\boldsymbol{q}} & -i2A_{\boldsymbol{q}} \end{pmatrix}\psi_{\boldsymbol{q}i}\equiv \tilde{H}_{\boldsymbol{q}}\psi_{\boldsymbol{q}i},
\end{equation}  
Comparing to the standard Kubo's theory \cite{PhysRevLett.105.225901}, the key ingredient of Qin's theory is an energy magnetization term: 
\begin{equation}
\kappa_{xy} = \kappa_{xy}^{\rm Kubo} + \frac{2M_E^z}{TV}.
\end{equation}
Here we take the phonon Hall conductivity in x-y plane as an example. $V$ is the volume in real space, and $T$ is the temperature. $M_E^z$ is the circulation of the phonon energy current named as the energy magnetization \cite{qin2011energy}. This correction term completely cancels the Kubo term to successfully avoid the divergence of the phonon Hall conductivity at zero temperature in the theory with Kubo term alone.
By solving the eigensystem, the Berry curvatures and phonon Hall conductivity are given by Qin $\emph{et~al.}$: 
\begin{equation}\label{eq:berry}
\boldsymbol{\Omega}_{\boldsymbol{q}i} =-\text{Im}\Big[\frac{\partial\bar{\psi}_{\boldsymbol{q}i}}{\partial\boldsymbol{q}}\times\frac{\partial\psi_{\boldsymbol{q}i}}{\partial\boldsymbol{q}}\Big],
\end{equation}
and 
\begin{equation} \label{eq:kappa}
\kappa_{xy}=-\frac{1}{2T}\int_{-\infty}^\infty d\epsilon\,\epsilon^2\sigma_{xy}(\epsilon)\frac{dn(\epsilon)}{d\epsilon},
\end{equation}
where
\begin{equation}
\begin{aligned}
&\bar{\psi} = \psi^{\dagger}\begin{pmatrix} D_{\boldsymbol{q}} &0\\ 0 &I\end{pmatrix},\\
&\sigma_{xy}(\epsilon)=-\frac{1}{V\hbar}\sum\limits_{\hbar\omega_{\boldsymbol{q}i}\le\epsilon}\Omega_{\boldsymbol{q}i}^z,
\end{aligned}
\end{equation}
$n(\epsilon) = 1/(e^{\epsilon/(k_BT)} - 1)$ is the Bose function at temperature
$T$,  $\epsilon$ represents the energy, and $k_B$ is the Boltzmann constant. The summation includes both positive and negative frequencies. The most common source of the $A_{\boldsymbol{q}}$ is the external magnetic field which has been applied in many experiments measuring the PHE. To describe this process, spin-phonon interaction was introduced. 

\subsection{Spin-phonon interaction}
After the first observation of the PHE in 2005, several researchers have tried to explain the experiments theoretically \cite{sheng2006theory, kagan2008anomalous, PhysRevLett.105.225901}, and all of them focused on the Raman-type spin-phonon interaction (SPI). Under an external magnetic field, the SPI in an ionic crystal lattice has the form of \cite{PhysRevLett.105.225901}
\begin{equation}
H_I = \sum\limits_i\boldsymbol{h}_{\alpha}\cdot (\boldsymbol{u}_{\alpha}\times \boldsymbol{p}_{\alpha})
\end{equation}
where $\boldsymbol{h}_{\alpha} = -\frac{q_{\alpha}}{2M_{\alpha}}\boldsymbol{B}$ if it is purely due to Lorentz force, $m_{\alpha}$ and  $q_{\alpha}$ are the ionic mass and charge at site ${\alpha}$, $\boldsymbol{u}_{\alpha}$ and $\boldsymbol{p}_{\alpha}$ are the vectors of displacement and momentum of the $\alpha$-th lattice site,  respectively. If one assumes the magnetic field is along z-axis, the SPI can be written as
\begin{equation}
H_I = u^TAp,
\end{equation}
where $A$ is an antisymmetric block diagonal matrix in real space with the diagonal block being $\begin{pmatrix} 0 &h_{\alpha}\\ -h_{\alpha} &0 \end{pmatrix}$. However, using $q_{\alpha}$ as the charge of the ion is not very accurate in real materials, and in fact, ionic materials do not have free charges. Instead, charge property should be described by a tensor, \emph{i.e.}, the Born effective charge tensor \cite{gonze1997dynamical}. With this correction, the $A$ matrix is:
\begin{equation}
\boldsymbol{A} = \frac{e}{4M_{\alpha}}(\boldsymbol{Z}_{\alpha}^T\times \boldsymbol{B} + \boldsymbol{B} \times \boldsymbol{Z}_{\alpha}),
\end{equation}
where $\boldsymbol{Z}_{\alpha}$ is the Born effective charge dyadic of the ion at site $\alpha$. The derivation and the meaning of the cross product are discussed in the appendix.

\subsection{An optimization: $\Theta(x)$}
Although equation (\ref{eq:kappa}) is enough to calculate the phonon Hall conductivity, it is usually difficult to implement the integral over the energy accurately if the Berry curvatures at some $\boldsymbol{q}$ points have large values. However, it is accessible to avoid this difficulty if we initially integrate over the energy by hand. In such a way, the formula of the phonon Hall conductivity becomes
\begin{equation}
\kappa_{xy} = \frac{k_B^2T}{2V\hbar}\sum\limits_{\boldsymbol{q},i}\Omega_{\boldsymbol{q}i}^z\Theta(\beta\hbar\omega_{\boldsymbol{q}i}),
\end{equation}
where $\beta = 1/k_BT$, and 
\begin{equation}
\Theta(x) = 
\begin{cases} 
\begin{aligned}\frac{x^2}{e^x-1}&-2x\,\ln(|e^x-1|)\\&+2{\rm Re}[{\rm Li}_2(e^{-x})]\end{aligned},&x\ne0,\\[20pt]
\pi^2/3,&x=0.
\end{cases}
\end{equation}
${\rm Li}_2$ is the so-called dilogarithm function. Although the dilogarithm function is still an integral, there are developed reliable packages to calculate it accurately in many languages such as Fortran, C++ and Mathematica. With this formula, the accuracy can be greatly boosted, therefore we call it an optimization. The details of the integration can be found in the appendix.

\section{Numerical details, results and discussion}
Dynamic matrix, vector potential, and Berry curvatures are the ingredients to calculate the phonon Hall conductivity. We determine the structures of the materials based on first-principles calculations using Quantum-Espresso (QE) \cite{QE-2009}, then calculate their interatomic force constants (IFC) up to the fourth-order with the help of the AIMD package in QE, and finally using the ALAMODE to extract the corresponding dynamic matrix including both analytic and non-analytic (with LO-TO splitting) part. We assume the vector potential is just from the SPI introduced in the last section with the block diagonal $A$ matrix. The Born effective charge dyadic is calculated by ph.x module in QE. As for the Berry curvatures, equation (\ref{eq:berry}) is too abstract to be used in a real calculation, but fortunately, converting it to a more explicit form using the eigen equation is already a common skill in topological physics. Taking the $z$-component of the Berry curvature as an example:
\begin{equation}\label{eq:explicitOmgea}
\Omega_{j,q_xq_y}^z = -{\rm Im}\Big[\sum\limits_{j\ne j^{\prime}}\frac{\bar{\psi}_j\frac{\partial\tilde{H}}{\partial q_x}\psi_{j^{\prime}}\bar{\psi}_{j^{\prime}}\frac{\partial\tilde{H}}{\partial q_y}\psi_j}{(\omega_j-\omega_{j^{\prime}} + i\eta)^2} - (q_x\leftrightarrow q_y)\Big],
\end{equation}    
where $\omega_j$ is the eigenfrequency in equation (\ref{eq:omega}), $\eta$ is related to the inverse of the phonon lifetime to avoid infinity when there are degenerate points. Since both the analytic and non-analytic part of the dynamic matrix have explicit formulas, and the SPI is independent of $\boldsymbol{q}$, the Berry curvatures can be explicitly worked out. Thereafter, the phonon Hall conductivity can be obtained by the summation of the weighted Berry curvatures in the first Brillouin zone.

\subsection{Numerical results for NaCl}
In 2011, Agarwalla \emph{et~al.} have calculated the PHE in NaCl using ``General Utility Lattice Program'' (GULP) with a Coulomb potential and a non-Coulomb Buckingham potential \cite{agarwalla2011phonon}. However, at that time, they used a not quite correct theory and their approach was still model-based. Therefore, we recalculate the PHE in NaCl in first-principles as a new benchmark. In our first-principles calculations, we apply structure optimization with the PAW-PBE pseudo-potential for Na and Cl to determine the lattice constant which turns out to be 5.65 \AA\,with the energy cutoff being 500 eV, and we use a $2\times2\times2$ supercell to calculate the IFCs. A $50\times50\times50$ grid and a $8\times8\times8$ grid are employed in calculating the dynamic matrix according to equation (\ref{eq:dynamic}) for $\boldsymbol{q}$ and $\boldsymbol{q^{\prime}}$ respectively. The small $\eta$ is chosen to be 0.1 ${\rm cm}^{-1}$.

\begin{figure}[h!]
\centering
\includegraphics[scale=0.28]{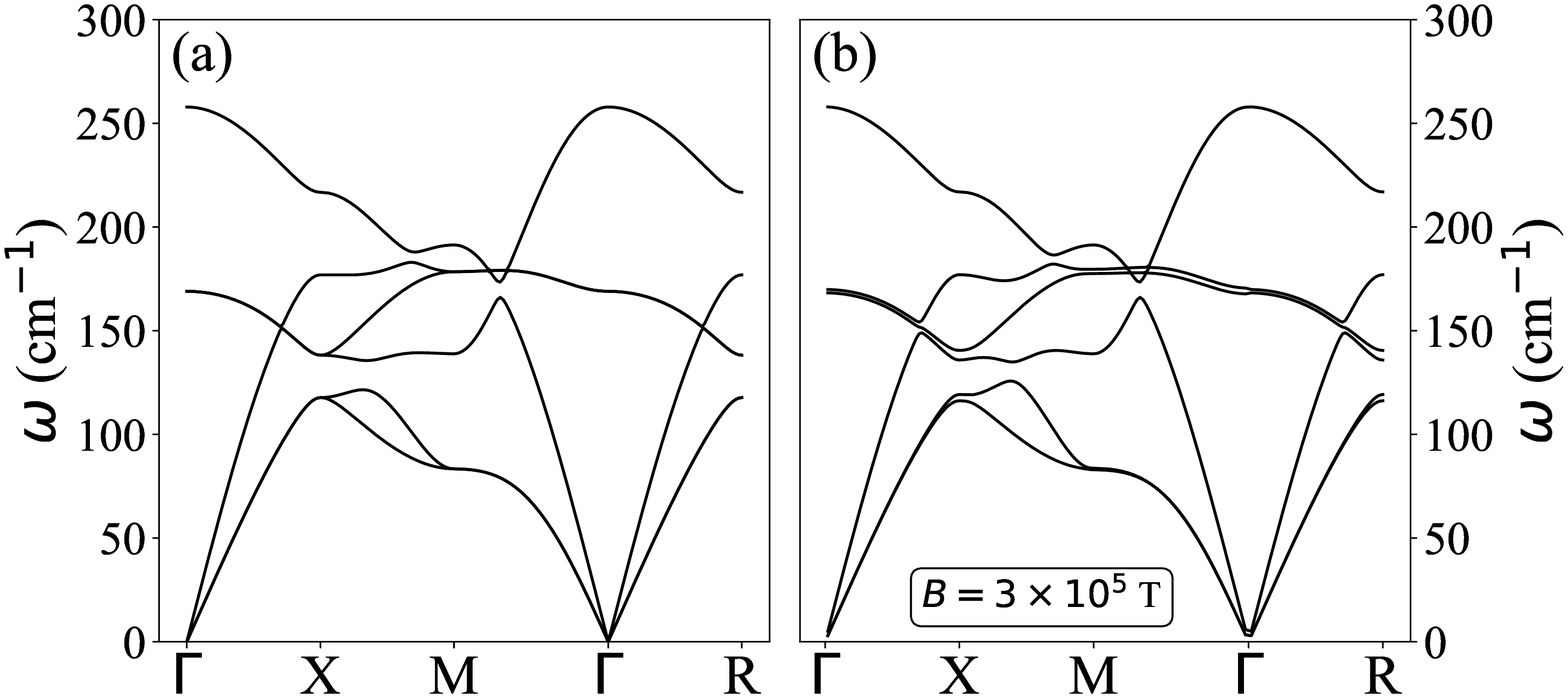}
\caption{(a) Phonon dispersion of NaCl at $T=300$ K without magnetic field. (b) Phonon dispersion of NaCl at $T=300$ K with an external magnetic field being $3\times 10^5$ T.} \label{fig:NaCl-dispersion}
\end{figure}

\begin{figure}[h!]
\centering
\includegraphics[scale=0.2]{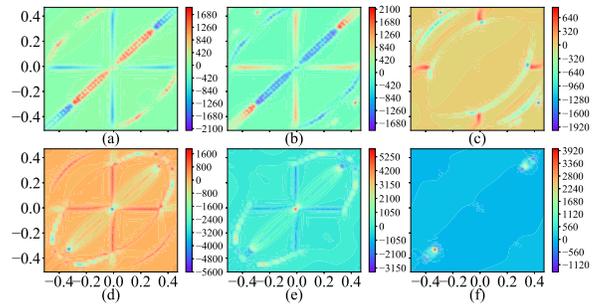}
\caption{The Berry curvatures of six positive branches in $\boldsymbol{b}_1-\boldsymbol{b}_2$ reciprocal plane of NaCl under the magnetic field $B=3\times 10^5$ T at temperature $T=300$ K, where $\boldsymbol{b}_1=\frac{2\pi}{a}(-\hat{q}_x+\hat{q}_y+\hat{q}_z),\,\boldsymbol{b}_2=\frac{2\pi}{a}(\hat{q}_x-\hat{q}_y+\hat{q}_z)$ are the two of three basis vectors with $a$ being the lattice constant. The horizontal and vertical axes represent the fraction of $\boldsymbol{b}_1$ and $\boldsymbol{b}_2$ in the range of $(-0.5, 0.5)$. The unit of the Berry curvatures is $a_0^2$, where $a_0$ is the Bohr radius. From (a) to (f), the associated eigenvalues are in ascending order.}
\label{fig:NaCl-BerryCurvatrues}
\end{figure}

Figure~\ref{fig:NaCl-dispersion}(a) shows the phonon dispersion of NaCl at $T=300$ K without an external magnetic field. LO-TO splitting is considered using the mixed-space approach \cite{wang2010mixed}. It can be seen that in Fig.~\ref{fig:NaCl-dispersion}(a) there are many degenerate points. If we apply a magnetic field (along z-direction throughout the paper) of $3\times 10^5$ T, those degenerate points will be lifted especially for the two TO modes as Figure~\ref{fig:NaCl-dispersion}(b) illustrates. Therefore, the role magnetic field plays is to open gaps in the phonon dispersion. Since NaCl has a simple structure, the branches in phonon dispersion can be well separated from each other by the applied magnetic field. As a result, we can draw the corresponding Berry curvatures of each branch, which are shown in Fig.~\ref{fig:NaCl-BerryCurvatrues}. Certain symmetries are observed in the Fig.~\ref{fig:NaCl-BerryCurvatrues}. The first and second acoustic branches are almost opposite to each other, so are the first and second optical branches, while the third acoustic branch and the third optical branch have their own patterns. This behaviour is consistent with the phonon dispersion of NaCl.

\begin{figure}[h!]
\centering
\includegraphics[scale=0.22]{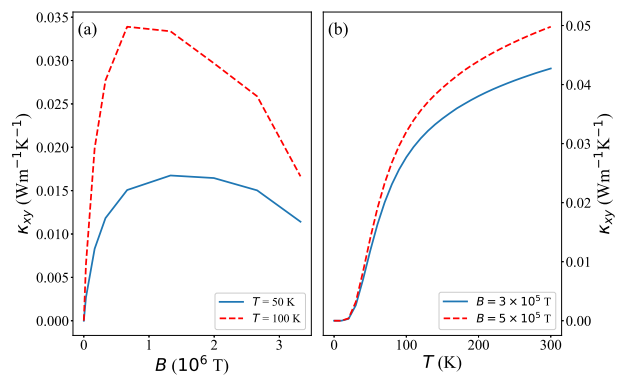}
\caption{(a) Phonon Hall conductivity versus the applied magnetic field at $T=50$ K and $T=100$ K respectively. (b) Phonon Hall conductivity versus temperature at $B=3\times 10^5$ T and $B=5\times 10^5$ T respectively.} \label{fig:NaCl}
\end{figure}

Figure~\ref{fig:NaCl} illustrates the dependence of the phonon Hall conductivity on magnetic field and temperature. It can be seen that as the temperature goes to 0, conductivity also decreases to 0. This is a favorable correction compared with the blowup of the conductivity near 0 K in Agarwalla \emph{et~al.}'s plots. For a small magnetic field, the magnitude of the conductivity is roughly linearly growing up, and when the magnetic field increases further, the magnitude starts to decrease, the same behavior as that in Agarwalla \emph{et~al.}'s results. However, the conductivity does not change sign in the same range of the magnetic field. Moreover, the magnitudes of our results are about one order larger than Agarwalla \emph{et~al.}'s, which is another progress of the \textit{ab initio} approach.

Although we obtain observable values of the phonon Hall conductivity, it requires a rather large magnetic field, about $10^5$ T at least. In experiments, a magnetic field with an order of magnitude 1 is enough to induce observable and even large phonon Hall conductivity in complex materials \cite{strohm2005phenomenological, li2020phonon}. Therefore, it deserves to implement our approach in some much more complicated materials such as materials in the family of perovskites.

\subsection{Numerical results for BTO}
BTO has a large dielectric constant, and it was argued that it is due to its soft optical phonons \cite{wang2020phonon} at $\Gamma$ point. Previous study implies that a large dielectric constant could result in large phonon Hall conductivity \cite{chen2020enhanced}, therefore, we calculate the PHE in BTO to verify this point. At different temperature ranges, BTO has different structures, while currently this structural diversity cannot be precisely caught by first-principles calculations \cite{evarestov2012first}. Therefore, we still choose the simple cubic BTO to implement the calculation. PAW-PBE pseudo-potentials for Ba, Ti, and O are employed with a $2\times2\times2$ supercell to calculate the dynamic matrix. The lattice constant is optimized to be 4.024 \AA, and the energy cutoff is set to be 800 eV. $\boldsymbol{q}$ and $\boldsymbol{q^{\prime}}$ grids are $50\times50\times50$ and $8\times8\times8$ respectively. The small $\eta$ is still chosen to be 0.1 ${\rm cm}^{-1}$.

\begin{figure}[h!]
\centering
\includegraphics[scale=0.3]{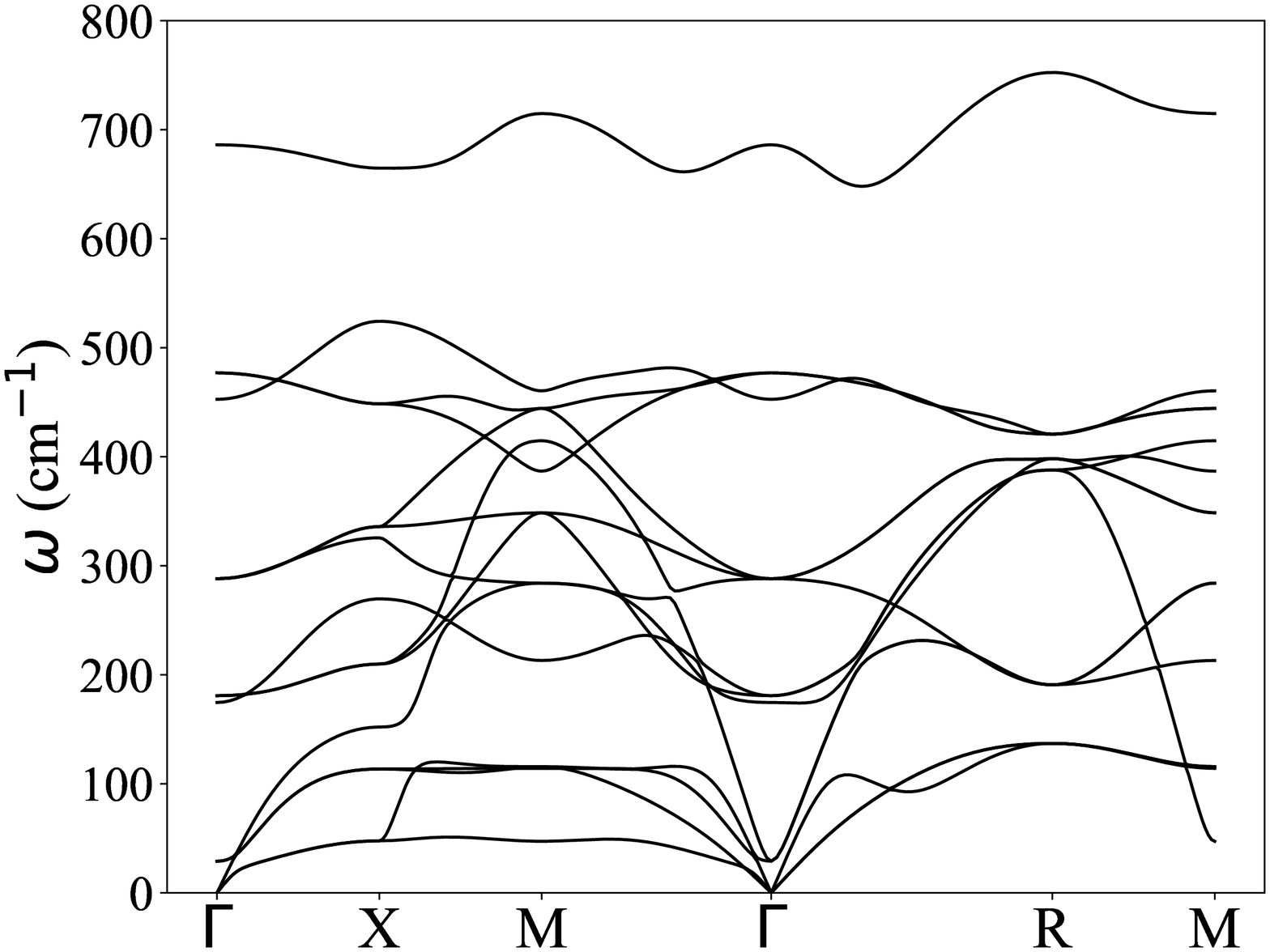}
\caption{Phonon dispersion of BTO at $T=60$ K without magnetic field.} \label{fig:BTO-dispersion}
\end{figure}

The phonon dispersion of BTO at $T=60$ K is illustrated in Fig.~\ref{fig:BTO-dispersion} where the two soft TO modes can be clearly seen near $\Gamma$ point whose frequencies are close to 0. Applying magnetic field results in a similar behavior as in NaCl which is trying to open gaps in dispersion. Since our goal for NaCl is to provide a benchmark while for BTO is to compare with experimental values, we use a reasonably large magnetic field with an order of magnitude 1 in this case. Within this range, the phonon dispersion almost remains the same under the magnetic field, therefore, it is not necessary to demonstrate it here.  

\begin{figure}[h!]
\centering
\includegraphics[scale=0.22]{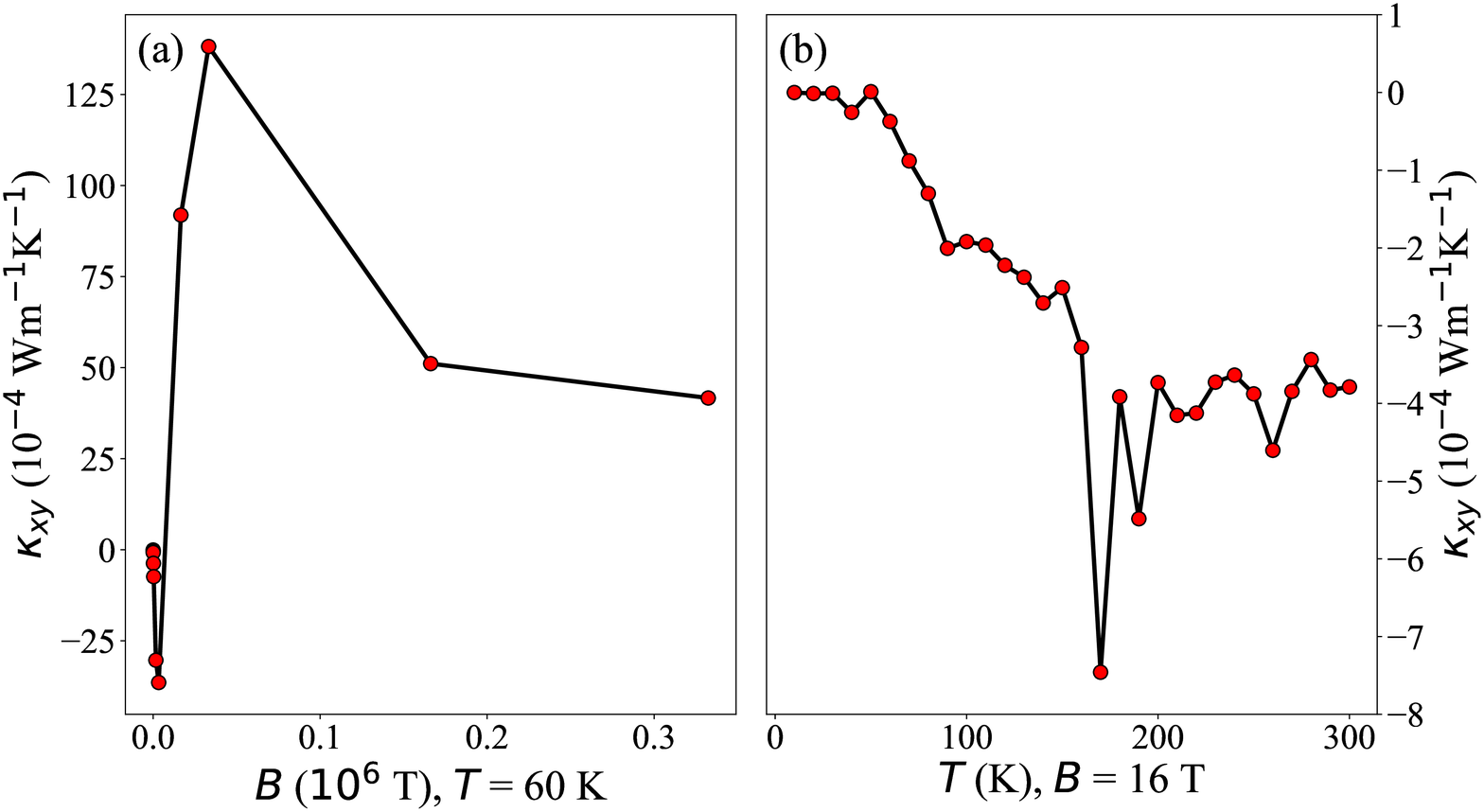}
\caption{(a) Phonon Hall conductivity versus the applied magnetic field at $T=60$ K. (b) Phonon Hall conductivity versus temperature at $B=16$ T.}\label{fig:BTO}
\end{figure}

Similar to Fig.~\ref{fig:NaCl}, Figure~\ref{fig:BTO} shows the behaviours of the phonon Hall conductivity against the magnetic field and temperature. Figure~\ref{fig:BTO}(a) is drawn at 60 K for this is roughly the lowest temperature range that first-principles calculations can correctly address the soft optical phonons in BTO \cite{wang2020phonon}. Again, for a small magnetic field, the Hall conductivity demonstrates a linear relationship with the magnetic field. For large fields, the phonon Hall conductivity also becomes large and even has a sign change. Figure~\ref{fig:BTO}(b) is under a magnetic field of 16 T, the absolute value of Hall conductivity increases at first and reaches a peak near 150 K, then starts to decrease. However, the order of magnitude is two orders smaller than the order of the experimental values in STO. Although STO and BTO are different materials, they have very similar crystal structures and both have soft optical modes at low temperatures \cite{he2020anharmonic}. Therefore, we think the comparison is reasonable. 

\begin{figure}[h!]
\centering
\includegraphics[scale=0.3]{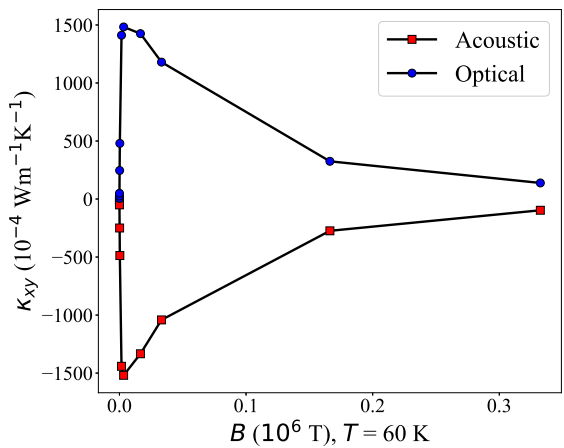}
\caption{Mode-dependent contributions to the phonon Hall effect for varying magnetic field at $T=60$ K. The red squares stand for the acoustic contributions and the blue dots for the optical contributions.}
\label{fig:BTO-mode-contribution}
\end{figure}

We note that when we enlarge the magnetic field, the phonon Hall conductivity in the BTO encounters a sign change. Since the conductivity is just the sum of the weighted Berry curvatures in the first Brillouin zone, we should observe clues for the sign change from the Berry curvatures and phonon dispersion of the BTO. Usually, the great change of Berry curvatures comes from band-openings or band-crossings. However, monitoring the evolution of each branch in the BTO is not a good idea. In the phonon dispersion of the BTO, many branches are deeply entangled so that we cannot always distinguish each branch correctly traveling around the whole Brillouin zone, neither the Berry curvatures of each branch. Moreover, the phonon Hall conductivity is an overall effect summing over all the weighted Berry curvatures so we cannot only analyze the individual Berry curvatures along the high symmetry path. Therefore, we decide to simply split the branches into two groups, three acoustic branches and twelve optical branches, and draw a plot of contributions to the phonon Hall conductivity of the two groups, which is the Fig.~\ref{fig:BTO-mode-contribution}. Comparing with the Fig.~\ref{fig:BTO}(a), we can conclude that the acoustic contributions are larger than optical for small magnetic fields so that the total conductivity is negative initially, and when the magnetic field surpasses some value, the situation gets reversed. Once a small magnetic field is applied to the system, the degenerate branches will be slightly lifted (points near the $\Gamma$ point are dominant) so that the Berry curvatures rapidly increase as shown in the Fig.~\ref{fig:BTO-mode-contribution}. Initial tiny gaps nearly produce symmetric Berry curvatures (dominated by the same $\eta$ in equation (\ref{eq:explicitOmgea}) ) among all the branches. However, due to the $\Theta$ function, the acoustic branches with much smaller eigenvalues will contribute more resulting in a negative conductivity (with a transformation, it is valid to just consider the positive branches \cite{PhysRevB.86.104305}). When the magnitude of the magnetic field keeps increasing, by zooming in the phonon dispersion, we find that the gaps in the acoustic branches grow faster than those in the optical branches against the magnetic field. As a result, the magnitude of the Berry curvatures of the acoustic branches decrease faster than those in the optical branches. The slopes of the two groups in the Fig.~\ref{fig:BTO-mode-contribution} verify this statement. Finally, at some value of the magnetic field, the optical branches contribute more to the phonon Hall conductivity so that a sign change shows up.

Why are the results so small? Our intuition is that the spin-phonon interaction, in this case, is too weak for it cannot even remove the degeneracy of the soft optical phonons. With this degeneracy, although we have soft optical phonons, their effects just get canceled. This canceling can be easily checked by looking at the mode contribution to the phonon Hall conductivity. However, currently we have no idea what are the suitable ingredients to open a gap between soft optical phonons from first-principles calculations, and we would like to leave it as an open question that deserves our further exploration. Therefore, we perform a numerical test to open a gap by hand. 

\begin{figure}[h!]
\centering
\includegraphics[scale=0.21]{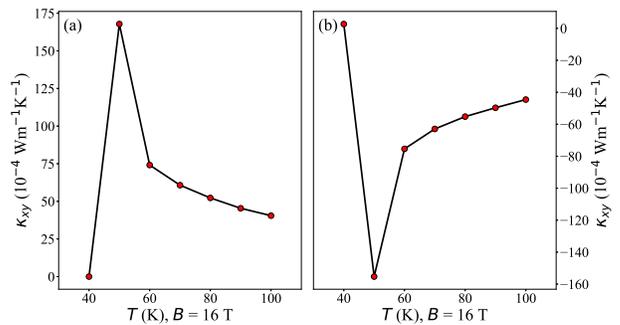}
\caption{(a) Open a small gap by manually lifting $1\%$ of the value of the higher soft optical phonon branch at and near $\Gamma$ point (the chosen range is where the frequencies are lower than 100 cm$^{-1}$). (b) Open a small gap with the same value and range as (a), but by manually lifting $1\%$ of the value of the lower soft optical phonon branch which will introduce band-crossing points near $\Gamma$ point. These two operations can be imagined considering a partially degenerate two-level system.} \label{fig:BTO-with-gap}
\end{figure}

There are two ways to manually open a small gap at and near the $\Gamma$ point, one is to lift the higher soft optical phonon branch and the other is to lift the lower soft optical phonon branch. The latter one will induce band-crossing points near the $\Gamma$ point. Figure~\ref{fig:BTO-with-gap} shows the Hall conductivity after these two operations. It can be seen that their magnitudes are indeed enlarged to be close to the experimental values. These two operations result in opposite signs, and usually the phonon Hall conductivity experiments measured has a negative sign.

\subsection{Discussion for STO}
Last year, an experimental group found a large phonon Hall conductivity in STO under the magnetic field around 15 T. Therefore, we also explored the PHE in STO by first-principles calculations. Since BTO and STO have a similar structure, the numerical details are almost the same as BTO except for the pseudo-potential files. Our optimized lattice constant for STO is 3.852 \AA\,based on the PBEsol exchange-correlation functional for Sr, Ti, and O \cite{PhysRevLett.100.136406}, which performs better than other functionals and is consistent with the previous experimental values \cite{okazaki1973lattice} and theoretical calculations \cite{tadano2015self}. However, we cannot obtain large phonon Hall conductivity even after manually open a gap, and the order of magnitude is still two orders smaller than the experiments in STO. The failure could result from many reasons. Firstly, we choose a cubic structure while at low temperature, STO has different phases of structure. Secondly, we expect there should be soft phonon modes with frequencies being close to 0 near $\Gamma$ point so that the dielectric constant of the STO will be as large as $10^4$ at low temperature, while our current approach utilizing ALAMODE cannot produce that soft optical modes, and the dielectric constant we obtained is about three orders smaller than expected. Thirdly, perhaps we cannot produce large PHE with the SPI. 

Right after the experiment, a theoretical paper by Chen \emph{et~al.} discussed this experiment in detail \cite{chen2020enhanced}. The authors pointed out that with Qin's theory, the phonon Hall conductivity can only be about four orders smaller than the experimental value. Although our results are two orders smaller, it is not large enough. Moreover, according to our observation, the SPI we used is too weak to open a gap between two soft phonon modes at $\Gamma$ point. Therefore the degeneration may cause canceling during the calculation. We obtain large values as those in the experiment if we open a gap by hand in BTO (not in STO for we cannot produce soft phonon modes in STO). In Chen \emph{et ~al.}'s paper, they also provide another direction to explain the experiment, which is using the Boltzmann transport theory. With their approach, they made a successful prediction of the ratio between the longitudinal conductivity and the phonon Hall conductivity. However, there is another new experiment in STO challenging their theory. Just by replacing the $^{16}$O in STO with the isotope $^{18}$O, researchers found that the phonon Hall conductivity will be reduced by two orders \cite{sim126sizable}. It is difficult to explain this behavior using Boltzmann transport theory for the replacement only changes the mass. Moreover, it is unnatural that we can only explain the PHE with macroscopic methods. 

When there is an external magnetic field, the ion will experience two effective vector potentials: one is from the real magnetic field (the SPI in our case), and the other is from the ``Berry phase'' due to the phase of electron ground state, which was first pointed out by Mead and Truhlar \cite{mead1979determination}. The latter one has already been considered in Qin's theory, and Saito \emph{et~al.} have discussed in detail how to include it in a square lattice model \cite{saito2019berry}. However, it seems that nobody knows how to calculate this electron-related vector potential in first-principles calculations. Another electron-related physical process is the spin-orbit coupling (SOC) of electrons. In our consideration, the SOC may affect the PHE in two ways. First, the SOC may relate to the electronic ``Berry phase'', but we cannot deal with it yet. Second, the SOC may modify the phonon dispersion directly. As a quick exploration, we calculate the phonon dispersion of the STO turning on the SOC at zero temperature, which is illustrated in the appendix. However, the effect of the SOC is rather weak that the phonon dispersion almost remains the same. Although previous research reported the SOC in the STO-based heterostructures \cite{kim2013origin} and gating system \cite{nakamura2012experimental}, there are no studies on the SOC in bulk STO or BTO before. Therefore, the high temperature effect of the SOC in the STO or BTO deserves future exploration. Besides, in our calculations, we do not take care of the cubic potential term which is related to the phonon lifetime. Qin's theory starts from the harmonic assumption, therefore we cannot deal with cubic term with this theory. Currently, we simply add a small constant value $\eta$ in equation (\ref{eq:explicitOmgea}) to represent the inverse of the phonon lifetime. Although we can tune the $\eta$ to modify the phonon Hall conductivity, a systematical theory for PHE considering the cubic term should be developed in future work. Therefore, we think the experiments still lack a microscopic explanation, and our intuition is that it may be relevant to the inner electronic topological structure of the STO or the cubic potential term in the STO, which is a future project to explore.

\section{Conclusion}
In summary, we introduce an approach to calculate the phonon Hall conductivity in real materials using first-principles calculations, and implemented it for NaCl, BTO, and STO. Although the approach is very direct, it highly relies on whether first-principles calculations can predict materials properly and how to introduce the effective vector potential in materials. We have provided a benchmark of the PHE in NaCl to be examined in the future, and based on our calculation, there is still a gap to address soft phonons in STO using first-principles calculations. We conclude that SPI is not a good candidate to explain the PHE in real materials, and propose that the inner electronic structure or cubic potential term in STO may be possible directions to explore in future work. Finally, we think the relationship between the soft mode and $\kappa_{xy}$ is far from clear quantitatively and needs further exploration. This study provides an effective route to capture the PHE from the accurate first-principles calculations in any real materials and has implications in promoting related experimental investigations.

\section*{Acknowledgments}
J.-S. W. is supported by a FRC grant R-144-000-402-114 and an MOE tier 2 grant
R-144-000-411-112. Z. GAO acknowledges the financial support from FRC tier 1 funding of Singapore (grant no. R-144-000-402-114). We also acknowledge the support by HPC Platform, Xi’an Jiaotong University.

\appendix
\section{Spin-phonon interaction with Born effective charge}
If we take a careful look at the form of SPI, it can be found that it has the similar form as the Hamiltonian containing a Lorentz force, therefore, to generalize it to couple with the Born effective charge, we should start from the magnetic energy. The energy of a magnetic moment is
\begin{equation}
V_m = -\boldsymbol{m}\cdot\boldsymbol{B},
\end{equation}
where usually $\boldsymbol{m}=\frac{e}{2}\boldsymbol{r}\times\boldsymbol{v}$, $e$ is the charge of the particle. Since the Born effective charge is a tensor, we should insert it into the equation carefully. A reasonable argument is from the way Born effective charge acting on the electric field, which is $\boldsymbol{Z}^T\boldsymbol{E}$. Here we take the transpose of $\boldsymbol{Z}$ because the first index of it is associated with the electric field \cite{gonze1997dynamical}. If we change the reference system so that the charge appears to move with a velocity $\boldsymbol{v}$, it will also feel a magnetic field $\boldsymbol{E}\rightarrow\boldsymbol{E} + \boldsymbol{v}\times\boldsymbol{B}$. Therefore, $\boldsymbol{Z}^T$ should act on $\boldsymbol{v}\times\boldsymbol{B}$, not on $\boldsymbol{B}$ directly. Moreover, in electronic system, the rate of change of the polarization is $\frac{d\boldsymbol{P}}{dt}=e\boldsymbol{Z}\boldsymbol{v}$. Analogous to this, we propose that in magnetic case, $\boldsymbol{Z}$ acts on $\boldsymbol{v}$. However, this replacement breaks the antisymmetry over $\boldsymbol{r}$ and $\boldsymbol{v}$. To restore it, we add a term with $\boldsymbol{Z}$ act also on $\boldsymbol{r}$ so that the energy becomes
\begin{equation}
\begin{aligned}
V_m &= -\frac{e}{4}[\boldsymbol{r}\times(\boldsymbol{Z}\boldsymbol{v}) + (\boldsymbol{Z}\boldsymbol{r})\times\boldsymbol{v}]\cdot\boldsymbol{B}\\
        &= -\frac{e}{4}[\big((\boldsymbol{v}\boldsymbol{Z}^T)\times\boldsymbol{B}\big)\cdot\boldsymbol{r} + (\boldsymbol{v}\times\boldsymbol{B})\cdot(\boldsymbol{Z}\boldsymbol{r})]\\
        &= -\frac{e}{4}[v_iZ_{ki}B_l\epsilon^{klj}r_j + \epsilon^{ikl}v_iB_kZ_{lj}r_j]\\
        &\equiv -\frac{e}{4}[\boldsymbol{v}\cdot(\boldsymbol{Z}^T\times\boldsymbol{B} + \boldsymbol{B}\times \boldsymbol{Z})\cdot\boldsymbol{r}]
\end{aligned}
\end{equation}
Then compare it with the form of the SPI, $H_I = u^TAp = -p^TAu$, we can conclude that
\begin{equation}
A = \frac{e}{4M_{\alpha}}(\boldsymbol{Z}_{\alpha}^T\times \boldsymbol{B} + \boldsymbol{B} \times \boldsymbol{Z}_{\alpha})
\end{equation} 

\section{The $\Theta$ function}
Given equation (\ref{eq:kappa}), we can firstly integrate with respect to energy.
\begin{equation}
\begin{aligned}
\kappa_{xy} &= \frac{1}{2TV\hbar}\sum\limits_{\boldsymbol{q},i}\Omega_{\boldsymbol{q}i}^z\int_{-\infty}^{\infty}d\epsilon\epsilon^2\theta(\epsilon-\hbar\omega_{\boldsymbol{q}i})\frac{dn(\epsilon)}{d\epsilon} \\
&\equiv\frac{k_B^2T}{2V\hbar}\sum\limits_{\boldsymbol{q},i}\Omega_{\boldsymbol{q}i}^z\Theta(\beta\hbar\omega_{\boldsymbol{q}i}),
\end{aligned}
\end{equation}
where $\theta$ is the step function, $n(\epsilon) = 1/(e^{\beta\epsilon} - 1)$ is the Bose function, $\beta=1/k_BT$, and
\begin{equation}
\Theta(x) = \int_{x}^\infty y^2dn(y),
\end{equation}
with the substitution $\beta\epsilon\rightarrow y$.
By integration by parts, we obtain
\begin{equation}
\Theta(x) = \frac{x^2}{e^x-1} + \int_{x}^{\infty}\frac{2ydy}{e^y-1}.
\end{equation}
When $x = 0$, the first term is an indeterminate value, but the original integral in this case has a definite value $\pi^2/3$. When $x \ne 0$, we make another substitution $y \rightarrow -\ln(u)$:
\begin{equation}
\Theta(x) = \frac{x^2}{e^x-1}-2\int_{0^+}^{e^{-x}}\frac{\ln(u)}{1-u}du.
\end{equation}
Again we use integration by parts,
\begin{equation}
\begin{aligned}
\Theta(x) & = \frac{x^2}{e^x-1} +2\ln(u)\,\ln(|1-u|)\Big|_{0^+}^{e^{-x}}\\
              &\qquad\qquad\,-2\int_{0^+}^{e^{-x}}\frac{\ln|1-u|}{u}du\\
& = \frac{x^2}{e^x-1}-2x\,\ln(|1-e^{-x}|)-2\int_{0^+}^{e^{-x}}\frac{{\rm ln}|1-u|}{u}du.
\end{aligned}
\end{equation}
The last term is related to the Spence's function or the dilogarithm function:
\begin{equation}
-\int_{0^+}^{x}\frac{\ln|1-u|}{u}du = \begin{cases} {\rm Li}_2(x), &x \le 1, \\[5pt]
\begin{aligned}\pi^2/3&-\ln^2(x)/2\\&-{\rm Li}_2(1/x)\end{aligned}, &x > 1.
\end{cases}
\end{equation} 
Since ${\rm Li}_2(x)+{\rm Li}_2(1/x) = \pi^2/6-\ln^2(-x)/2$, we can combine two cases so that finally we obtain

\begin{equation}
\Theta(x) = 
\begin{cases} 
\begin{aligned}\frac{x^2}{e^x-1}&-2x\,\ln(|e^x-1|)\\&+2{\rm Re}[{\rm Li}_2(e^{-x})]\end{aligned}, &x \ne 0, \\[20pt]
\pi^2/3, &x = 0.
\end{cases}
\end{equation}
Here we always take the real part of the ${\rm Li}_2(e^{-x})$ for when $e^{-x} > 1$, it is a complex value while $\Theta(x)$ is real.

\section{Spin-orbit coupling in STO}
\begin{figure}[h!]
\centering
\includegraphics[scale=0.3]{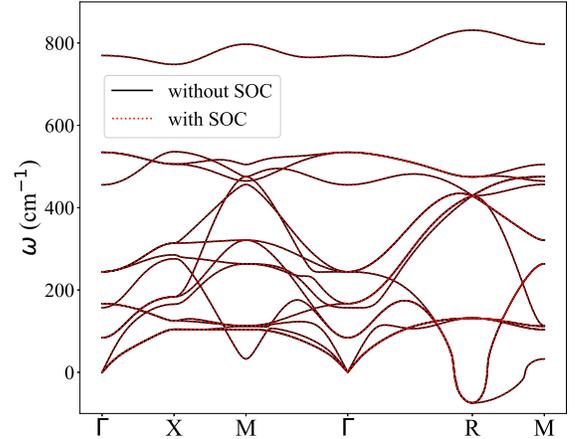}
\caption{The phonon dispersion of the STO with and without the SOC at 0 K. The black solid line stands for the case without the SOC, and the red dotted line for the case with the SOC.}
\label{fig:noSCO-SCO-STO}
\end{figure}
We calculate the phonon dispersion of the STO with the SOC considered at zero temperature. The numerical details are the same as we introduced in the main text except the SOC turned on during the first-principles calculations. The comparison is given in the Fig. \ref{fig:noSCO-SCO-STO}. It can be seen that the effect of the SOC at zero temperature is too weak to modify the phonon dispersion obviously. The effect of the SOC at higher temperature has not been reported yet, and currently, precise first-principles calculations for STO at non-zero temperatures are still challenging.
\vfill  

\bibliography{references}

\end{document}